\documentclass[10pt, conference]{IEEEtran}

\usepackage{pifont}
\usepackage{amsmath}
\usepackage{wasysym}

\usepackage[colorlinks = true,
  linkcolor = blue,
  citecolor = blue,
  urlcolor = blue]{hyperref}

\usepackage{placeins}
\usepackage{graphicx}
\usepackage{booktabs}
\usepackage{multirow}
\usepackage{tabularx}
\usepackage{balance}
\usepackage{url}

\newcommand{\fullcircle}{\CIRCLE}
\newcommand{\whitecircle}{\Circle}
\newcommand{\partialcircle}{\LEFTcircle}
\newcommand{\onecase}{\ding{192}}
\newcommand{\twocase}{\ding{193}}
\newcommand{\one}{\ding{182}}
\newcommand{\two}{\ding{183}}
\newcommand{\three}{\ding{184}}
\newcommand{\four}{\ding{185}}

\usepackage{fancyhdr} 
\fancypagestyle{firststyle}
{
\fancyhf{}
\fancyfoot[C]{\scriptsize{Proceedings of the IEEE International Conference on Software Analysis, Evolution and Reengineering -- Companion (SANER-C 2026), Limassol, IEEE, pp. 325--332. \\ This version is the authors' copy. The publisher's definite version is available online via \url{https://doi.org/10.1109/SANER-C67878.2026.00050}.}}}

\begin{document}

\title{Empirical Derivations from an Evolving Test Suite}

\author{
\IEEEauthorblockN{Jukka Ruohonen}
\IEEEauthorblockA{University of Southern Denmark \\
Email: juk@mmmi.sdu.dk}
\and
\IEEEauthorblockN{Abhishek Tiwari}
\IEEEauthorblockA{University of Southern Denmark \\
Email: abti@mmmi.sdu.dk}
}

\maketitle

\begin{abstract}
The paper presents a longitudinal empirical analysis of the automated,
continuous, and virtualization-based software test suite of the NetBSD operating
system. The longitudinal period observed spans from the initial roll out of the
test suite in the early 2010s to late 2025. According to the results, the test
suite has grown continuously, currently covering over ten thousand individual
test cases. Failed test cases exhibit overall stability, although there have
been shorter periods marked with more frequent failures. A similar observation
applies to build failures, failures of the test suite to complete, and
installation failures, all of which are also captured by the NetBSD's testing
framework. Finally, code churn and kernel modifications do not provide
longitudinally consistent statistical explanations for the failures. Although
some periods exhibit larger effects, including particularly with respect to the
kernel modifications, the effects are small on average. Even though only in an
exploratory manner, these empirical observations contribute to efforts to draw
conclusions from large-scale and evolving software test suites.
\end{abstract}


\section{Introduction}

\thispagestyle{firststyle} 

Operating systems require and benefit from testing like any other
software. However, continuous and automatic testing of operating systems has
long been a challenge due to the nature of operating systems, including their
kernels who deal with hardware and low-level characteristics in general. Despite
the overall challenge, progress has been made during the past two decades,
including in the open source software (OSS) world. Continuous and automated
fuzzing of OSS operating system kernels would be a good
example~\cite{Ruohonen19RSDA}. Also more traditional automated testing
frameworks suites have been introduced.

The NetBSD's test suite is based on automated testing framework developed in the
early 2010s. Among other things, the framework's benefits include automated runs
in a virtual machine~\cite{Kantee10}. Specifically, the NetBSD project has a
continuous system that continuously builds the whole operating system from
source code, install the whole operating system to a Qemu-based virtual machine,
and runs the individual test cases within the virtual machine installation. The
individual test cases cover anything from unit to integration and stress tests.

In addition, the testing frameworks supports different instruction set
architectures as well as the NetBSD's own implementations for running kernel
code in userland~\cite{Husemann11}. All in all, the testing framework can be
seen as a NetBSD's answer to the common problem about a lack of a universal
automation tool for testing operating systems, including those used for embedded
devices~\cite{AliShah16}. Given these introductory points, the test suite
provides an excellent case study for examining what can be empirically and
longitudinally derived from the test results outputted by the continuous testing
framework. To the best of the author's knowledge, no prior work has been done to
examine the NetBSD's test suite. Regarding the phrasing about what can be
derived, the opening Section~\ref{sec: background and related work} presents
three research questions (RQs) for motivating the empirical derivation in
relation to existing research. Materials and methods are subsequently described
in Section~\ref{sec: materials and methods}. Results follow in the next
Section~\ref{sec: results}. A conclusions is presented in Section~\ref{sec:
  conclusion}. Sections~\ref{sec: limitations} and \ref{sec: further research}
present brief takes on limitations and further research possibilities,
respectively.

\section{Motivation and Related Work}\label{sec: background and related work}

\subsection{Software Evolution}

The so-called ``laws'' of software evolution have been continuously discussed,
theorized, and evaluated during the past three or four decades of software
engineering research~\cite{Herraiz13}. Although originally these were formulated
rather vaguely, thus leaving a wide room for interpretation, the laws' basic
message resonates with the discipline's core premises; the sizes of most
software projects grow continuously, and, therefore, also complexity is
continuously increasing and must be dealt with by continuous modifications,
whether maintenance measures or refactoring~\cite{Ruohonen15JSEP}. The later
metaphor of technical debt conveys the same message~\cite{Siebra14}. For
robustly dealing with the accumulating debt, also a test suite should grow
continuously; otherwise, refactoring and other maintenance measures may be
risky. Although test coverage is not observed, the following simple question is
thus fundamental and hence worth asking:

\begin{itemize}
\item{RQ.1: \textit{Has the number of individual test cases in the NetBSD test suite continuously increased over time?}}
\end{itemize}

The second research question is about failure trends:

\begin{itemize}
\item{RQ.2: \textit{Have the amounts of \one~failed test cases, \two~build
    failures, \three~failures of the test suite to complete, and
    \four~installation failures remained stable over the years?}}
\end{itemize}

Answers to RQ.2 can be used to deduce, even if only tentatively, whether the
testing results indicate periods during which the NetBSD project might have had
development and maintenance problems. With respect to RQ.2\one, a growth trend
instead of longitudinal stability could be seen as a sign that some problems are
present. Alternatively, large amounts of failed test cases, or other failures
indicated by the test suite, in some particular period could signpost
intensified or large-scale restructuring and refactoring or the introduction of
large features during which failures are presumably more common. Although with
negative results, a somewhat similar reasoning has been previously used in
relation to major FreeBSD releases and volatility of development activity
metrics~\cite{Ruohonen15IWPSE}. This reasoning is worth pointing out because
FreeBSD has traditionally relied on feature testing and manual testing of
releases before their launches~\cite{DinhTrong04}. With this point in mind, the
NetBSD's testing framework has also improved release
engineering~\cite{Husemann11}. Thus, all in all, also RQ.2 can be linked to
existing research.

\subsection{Code Churn and Kernel Code}\label{subsec: code churn and kernel code}

The third research question seeks to probe whether the longitudinal failure
amounts correlate with two software metrics:

\begin{itemize}
\item{RQ.3: \textit{Can \onecase~code churn and \twocase~kernel modifications
    prior to test runs explain the \one~amounts of failed test cases,
    \two~whether a build failed, \three~whether the test suite failed to
    complete, and \four~whether an installation failed?}}
\end{itemize}

With respect to RQ.3\onecase, there is a vast literature branch using different
code churn metrics for modeling and predicting different software engineering
phenomena. The application domains range from bug~\cite{Giger11} and
vulnerability~\cite{Theisen20} prediction to
fuzzing~\cite{Ruohonen19RSDA}. Given these domains, it seems sensible to
hypothesize that increased churn would also increase the amounts of test case
failures and other failures detected by the NetBSD's automated test
suite. Regarding RQ.3\twocase, the motivation is even more on the exploratory
side. Nevertheless, there are some existing results hinting that kernel
modifications could correlate with some particular failure types. For instance,
it seems sensible to assume that introducing bugs to kernel code would increase
particularly a likelihood that a subsequent installation fails. Empirically, the
assumption makes sense because the majority of code changes in operating system
projects are either about userland code or kernel code, including in the NetBSD
operating system~\cite{Kudrjavets22}. Thus, also RQ.3 can be connected to
existing research albeit only loosely.

Despite the three RQs and their brief motivations, it should be emphasized that
the paper is still an exploratory case study whose ``primary purpose is to
examine a little understood issue or phenomenon to develop preliminary ideas and
move toward refined research questions by focusing on the
`what'~question''~\cite[p.~53]{Kempenes08}. Although there are existing
empirical studies on the evolution of test suites~\cite{Aljedaani20, Pinto12},
as said, the NetBSD's test suite has not been previously examined. The knowledge
seems limited also regarding test suites for operating systems in general. With
respect to the phrasing about ``little understood issue or phenomenon'' in the
previous quote, RQ.3 is the reference point. To best of the author's knowledge,
the question has neither been examined nor answered previously. Thus, even the
``preliminary ideas'' noted in the quote can provide insights for
moving~forward.

\section{Materials and Methods}\label{sec: materials and methods}

\subsection{Data}

The dataset was collected from the NetBSD's online archives for the results from
the automated test runs~\cite{NetBSD25a}. The dataset is also available online
for replication and other purposes~\cite{Ruohonen25DSNETBSD}. The period
observed ranges from the first day of August 2011 to the last day of September
2025. Although NetBSD supports multiple instruction set
architectures~\cite{NetBSD25b}, the analysis is restricted to the conventional
32-bit and \text{64-bit} \texttt{x86} architectures, as denoted in NetBSD with
the \texttt{i386} and \texttt{amd64} labels. The reason is that the test suite
has successfully ran the longest for these two architectures. Although also
other architectures are tested with the automated framework, the testing has
started later than with \texttt{i386} and \texttt{amd64}. Furthermore, some of
the more exotic architectures have occasionally been broken for relatively long
periods of time, meaning that an installation has failed and thus also the test
suite has failed to execute. Such periods cause irregularities and other
disturbances for the corresponding time series.

In practice, the dataset was assembled by parsing each file in the online
archives line by line. As soon further discussed in Subsection~\ref{subsec:
  aggregation}, the archival files are provided for each architecture on monthly
basis. The following snippet~\cite{NetBSD25c} can be used to illustrate the
nature of the textual archival data:

\begin{scriptsize}
\begin{verbatim}
  build: OK with 737188 lines of log, install: OK,
         tests: 10226 passed, 608 skipped,
         87 expected_failure, 2 failed,
         ATF output: raw, xml, html
         new failures: lib/libc/gen/t_sleep/kevent
  commit 2025.01.23.12.32.38 christos
         src/tests/lib/libexecinfo/Makefile 1.9
  commit 2025.01.23.12.32.38 christos
         src/tests/lib/libexecinfo/t_backtrace_sandbox.c 1.1
  build: failed with 646199 lines of log
  commit 2025.01.23.12.36.14 christos
         src/distrib/sets/lists/debug/mi 1.460
  commit 2025.01.23.12.36.15 christos
         src/distrib/sets/lists/tests/mi 1.1358
  build: OK with 737095 lines of log, install: OK,
         tests: 10226 passed, 608 skipped,
         87 expected_failure, 3 failed [...]
\end{verbatim}
\end{scriptsize}

When starting from the top, it can be seen that a test suite run completed with
over ten thousand passed tests. Afterwards, two commits were made to files
within the \texttt{src/tests} directory where the tests reside. These broke the
build. After two subsequent commits were made, the build again succeeded, but
the corresponding test suite run indicated one additional test failure compared
to the previous successful run. Although only a build failure is shown in the
snippet, the test suite also records a failure in case a Qemu-based installation
failed. If a build and an installation succeeded, but the test suite failed to
run successfully in the virtual machine installation, a further failure is
recorded. These failure cases contain everything and anything from timeouts to
hangs of the operating system itself.

\subsection{Metrics}\label{subsec: metrics}

The NetBSD's test suite classifies test results into four categories: passed,
skipped, failed, and expectedly failed. The skipped category is used for
so-called ``flaky tests''---tests whose results are
non-deterministic~\cite{Pinto20}, as well as for other related reasons. The
expected failure category denotes test cases in which a developer has explicitly
flagged a failure as an expected result. These cases range from long-lasting
bugs to other known outcomes, such as features in standards that have not been
implemented. The failed category is used for RQ.2, while an answer to RQ.1 is
solicited simply by summing all the outcome categories and then carrying out an
aggregation. The three failure cases noted in the preceding subsection provide
the remaining metrics for answering to the question~RQ.2. Regarding metrics used
in the domain of software testing and their classifications more
generally~\cite{Witte24}, the metrics used are about testing and test case
quantities. This characterization is further reinforced by the aggregation
soon~discussed.

Although no universally accepted definitions are available for code churn, it
has typically been measured as source code lines changed, deleted, and added
between commits or larger aggregates such as releases~\cite{Ruohonen19RSDA,
  Giger11, Kudrjavets22}. The present work diverges from this tradition by using
plain commit counts as a proxy for churn, as has been done also in previous
research~\cite{Kudrjavets23}. With RQ.3\onecase~and RQ.3\two~in mind, the
example snippet shown earlier would yield a value two in relation to the
particular build failure shown in the snippet.

Regarding the kernel modifications, these were identified by commits involving
files in the \texttt{sys} root directory of the NetBSD's source code tree. Also
this operationalization has been previously used~\cite{Kudrjavets06}. Given that
these kernel commits and the total commit counts correlate, the answer to
RQ.3\twocase~is sought with a dummy variable taking a value zero unless any of
the preceding commits before a test run touched files in the \texttt{sys}
directory in which case the variable's value is one.

\subsection{Aggregation and Units of Analysis}\label{subsec: aggregation}

For RQ.1 and RQ.2, monthly aggregates are used, meaning that the length of the
time series is $170$. This aggregation frequency is a natural choice in a sense
that NetBSD provides separate monthly data archives about the testing
results. Regarding aggregation functions, mean and median are used to the
amounts of tests and failed tests. For the build failures, test suite failures,
and installation failures, monthly sums are used because averaging is not a good
choice for these metrics. In other words, a count of build failures in a month,
for instance, conveys better information about potential problems than some
average amount of build failures recorded during the month.

It is important to further emphasize that RQ.1 and RQ.2\one~are approached by
deleting the observations in which a build, the test suite, or an installation
failed prior to the aggregation operations. Without the deletion, there would be
undefined values in these two two series; no failed test cases, or any other
testing results, are recorded under the three failure cases. However, a similar
deletion does not obviously make sense for seeking answers to RQ.2\two,
RQ.2\three, and RQ.2\four.

\begin{figure}[th!b]
\centering
\includegraphics[width=\linewidth, height=3cm]{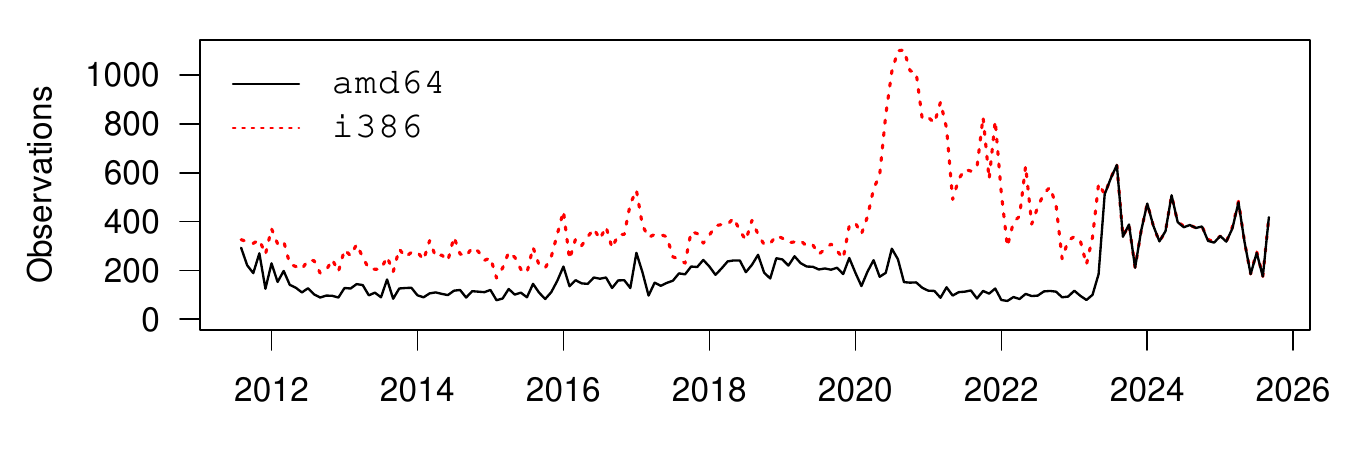}
\caption{Monthly Sample Sizes}
\label{fig: monthly n}
\end{figure}

Regarding RQ.3, the regression analyses soon described in
Subsection~\ref{subsec: methods} are computed primarily with unaggregated
monthly data. The reason is partially practical and related to the data
collection: because NetBSD provides the per-architecture online archival files
on rolling monthly basis, it is not straightforward to merge and robustly
synchronize the files into annual aggregates or a single large dataset covering
the whole period observed. Furthermore, it should be emphasized that the monthly
archival files and hence also the monthly unaggregated samples vary across the
architectures. The reason is that the virtualization-based test suite is
executed at different paces for different architectures. Although the
documentation, which may be outdated, says that the test suite is being run
circa once per day for \texttt{amd64} and circa six times per day for the
\texttt{i386} architecture~\cite{NetBSD25a}, Fig.~\ref{fig: monthly n} indicates
that this scheduling seems to have applied only between circa 2021 and 2024. In
other words, when this period is excluded, the sample sizes between
\texttt{i386} and \texttt{amd64} are roughly comparable.

\begin{figure*}[t!]
\centering
\includegraphics[width=\linewidth, height=6cm]{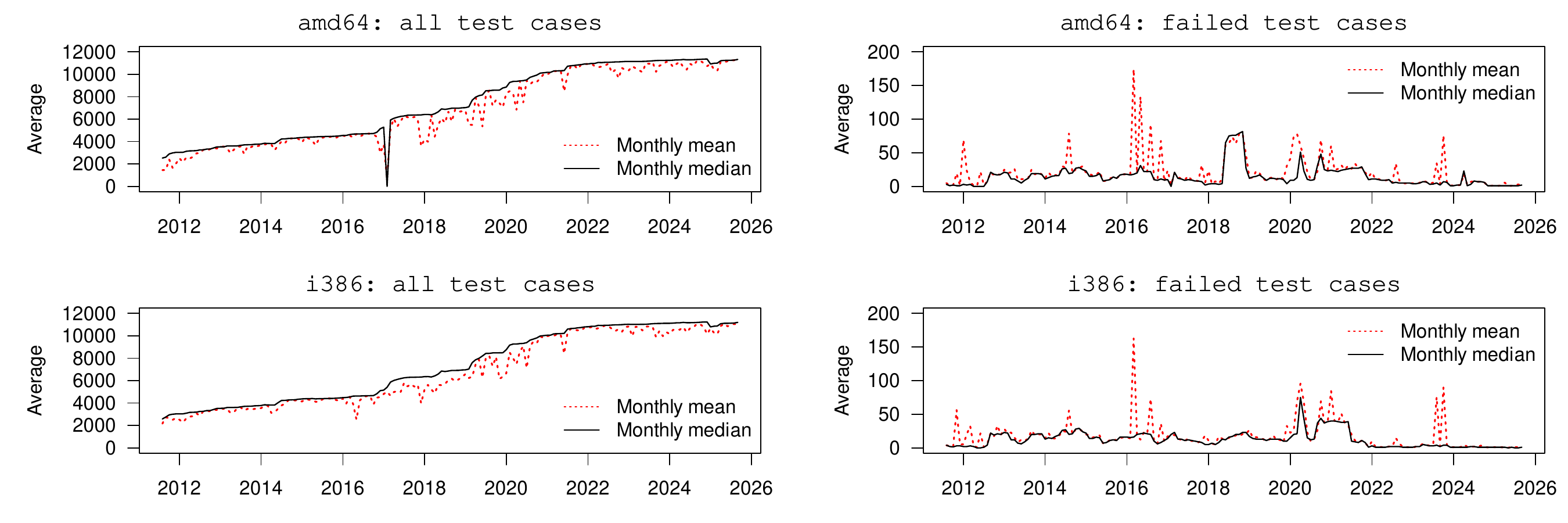}
\caption{Test Cases and Failed Test Cases (monthly averages)}
\label{fig: tests}
\end{figure*}

\subsection{Methods}\label{subsec: methods}

Methods were selected based on data rather than having been fixed in
advance. That is no, no pre-registration was done. Thus, a simple visual
inspection is sufficient for answering to the first research question
RQ.1. Although formal time series methods and tests, such as those originating
from fluctuation processes~\cite{Ruohonen25ICTSSa}, could be used for answering
to RQ.2, a visual interpretation provides the primary means for answering also
to this research question. By following recent research~\cite{Ruohonen25CLSR},
the Kruskal-Wallis~\cite{Kruskal52} is briefly also used to check whether the
failures differ between two periods from August 2011 to August 2018 and
September 2018 to September 2025.

The ordinary least squares (OLS) regression is used for probing a tentative
answer to RQ.3\one~and logistic regression~(LR) to the remaining RQ.3\two,
RQ.3\three, and RQ.3\four. Code churn, as measured by the commit counts, and the
dummy variable for kernel modifications are the only two explanatory metrics
used. Therefore, the interest is \textit{not} to seek good overall statistical
performance but to check whether the two metrics correlate with failure
amounts. Regarding the two explanatory metrics, these refer to preceding
observations prior to a given test suite run or a failure; the counting starts
over from zero once the given run or failure has occurred in the archival files.

As noted in the preceding Subsection~\ref{subsec: aggregation}, unaggregated
monthly data is used for both the OLS and LR regressions, meaning that in total
$4 \times 2 \times 170 = 1,360$ individual regressions are estimated, given the
four different dependent variables, two architectures, and the length of the
time series. Therefore, a healthy grain of salt should be taken particularly
with statistical significance. Among other things, both model calibration and
diagnostics are difficult to do with such a large amount of regressions, a
Bonferroni correction might be considered, and so forth and so on. For these
reasons, the results are reported by focusing on the regression coefficients
from the OLS regressions and the marginal effects from the LR regressions; the
latter proxy affects directly upon probabilities. Even with the limitations
noted, the monthly regressions have a benefit in that these can reveal whether
effects are present in some particular months or longer periods. That is: even
in case the effects are modest generally, it could be that some particular
months or periods still show strong statistical effects.

\begin{figure}[t!]
\centering
\includegraphics[width=\linewidth, height=9cm]{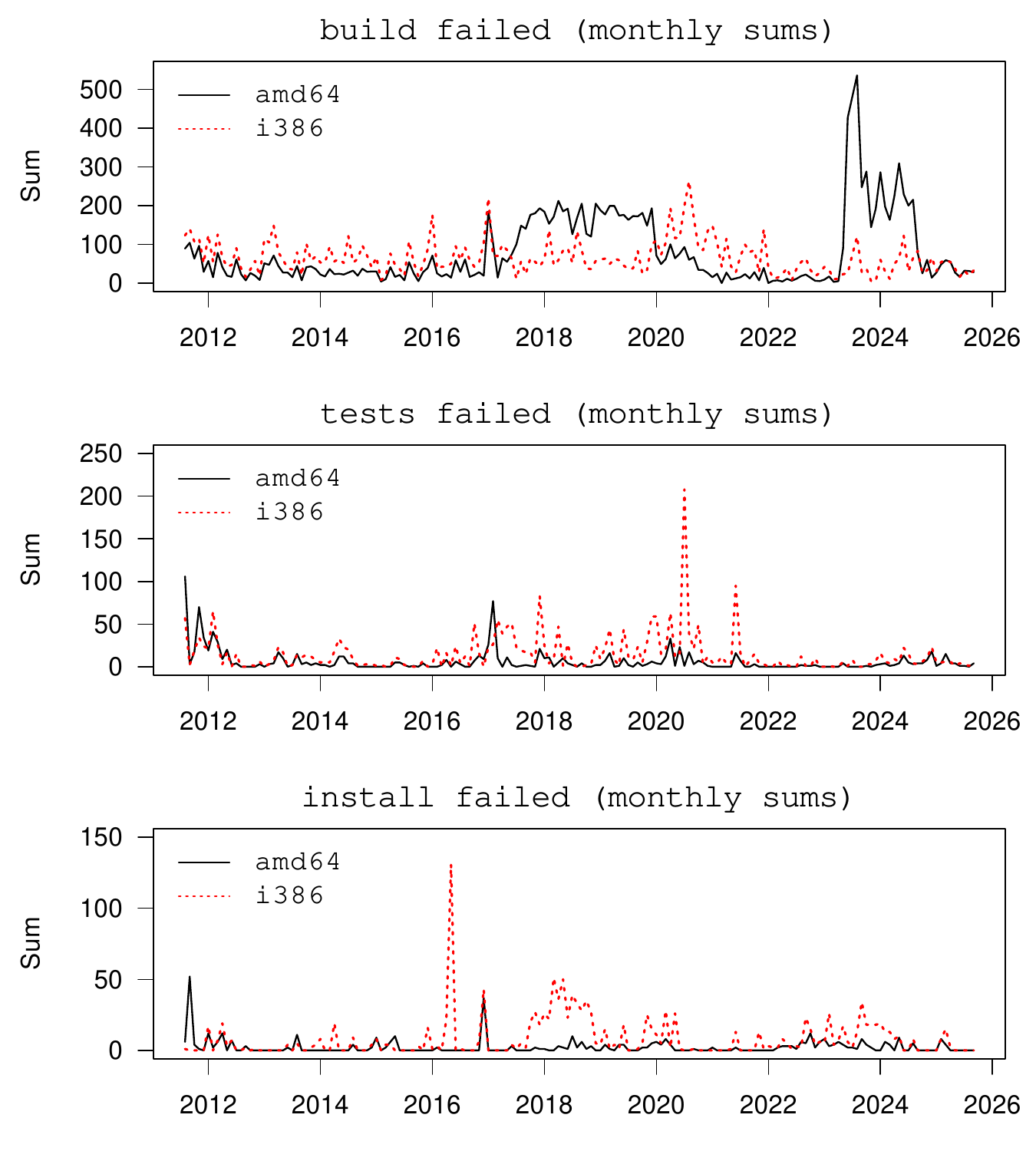}
\caption{Build, Test Suite, and Installation Failures (monthly sums)}
\label{fig: failures}
\end{figure}

In addition, brief checks are computed with pooled unaggregated data for the two
architectures. The amount of observations are $65,001$ and $31,840$ for
\texttt{i386} and \texttt{amd64}, respectively. Besides OLS and LR regressions
with the pooled unaggregated data, the results for the failed test cases are
also checked by estimating a negative binomial regression with a log-link and a
first-order autoregression, AR(1), term using an existing
implementation~\cite{Liboschik17}. The reason for this additional check is that
the pooled time series exhibit pronounced count data characteristics. However,
as was noted in Subsection~\ref{subsec: aggregation}, the pooled estimates
suffer from a limitation that the monthly archival files are not properly
synchronized. Therefore, a grain of salt is again needed for interpretation.

\section{Results}\label{sec: results}

\subsection{Growth (RQ.1)}\label{subsec: growth}

The answer to RQ.1 is clear: the test suite has continuously grown---as has most
OSS operating systems themselves ~\cite{Zhao25}, including NetBSD presumably
with them. The answer also places the NetBSD operating system into a distinct
cluster of OSS projects with growing test suites~\cite{Miranda25}. This
observation can be seen from the left-hand side plot in Fig.~\ref{fig: tests},
which shows the monthly means and medians of the individual test case
amounts. The last observations in the dataset indicate as many as $11,326$ test
cases for \texttt{amd64} and $11,198$ for the old \texttt{i386}
architecture. Even when keeping in mind that operating systems, including
NetBSD, have large code basis, these amounts are large enough to conclude that
the automated test framework has extensively and continuously been used in
NetBSD throughout the period observed. However, the amount of new test cases
added seems to have slowed down or stabilized from circa 2022 onward. Although
testing coverage is not observed, as already noted, this observation could be
used to contemplate in further research whether a satisfactory coverage level
has been reached. Alternatively, it could also be that most of what can be
easily tested is already tested.

\begin{figure*}[th!b]
\centering
\includegraphics[width=\linewidth, height=12.8cm]{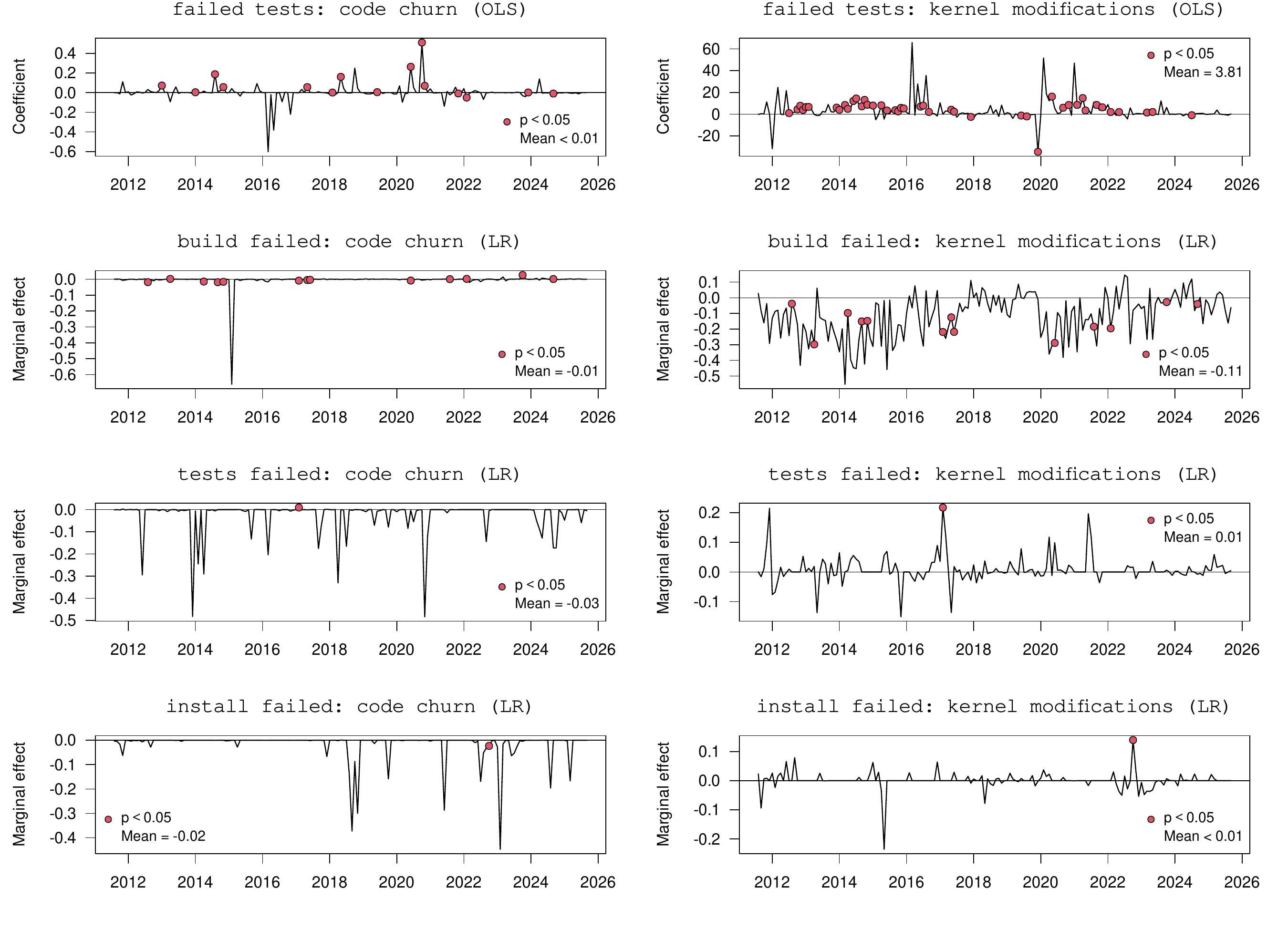}
\caption{Regression Results for the \texttt{amd64} Architecture (separate monthly estimates; coefficients or marginal effects of the coefficients)}
\label{fig: reg amd64}
\end{figure*}

\begin{figure*}[th!b]
\centering
\includegraphics[width=\linewidth, height=12.8cm]{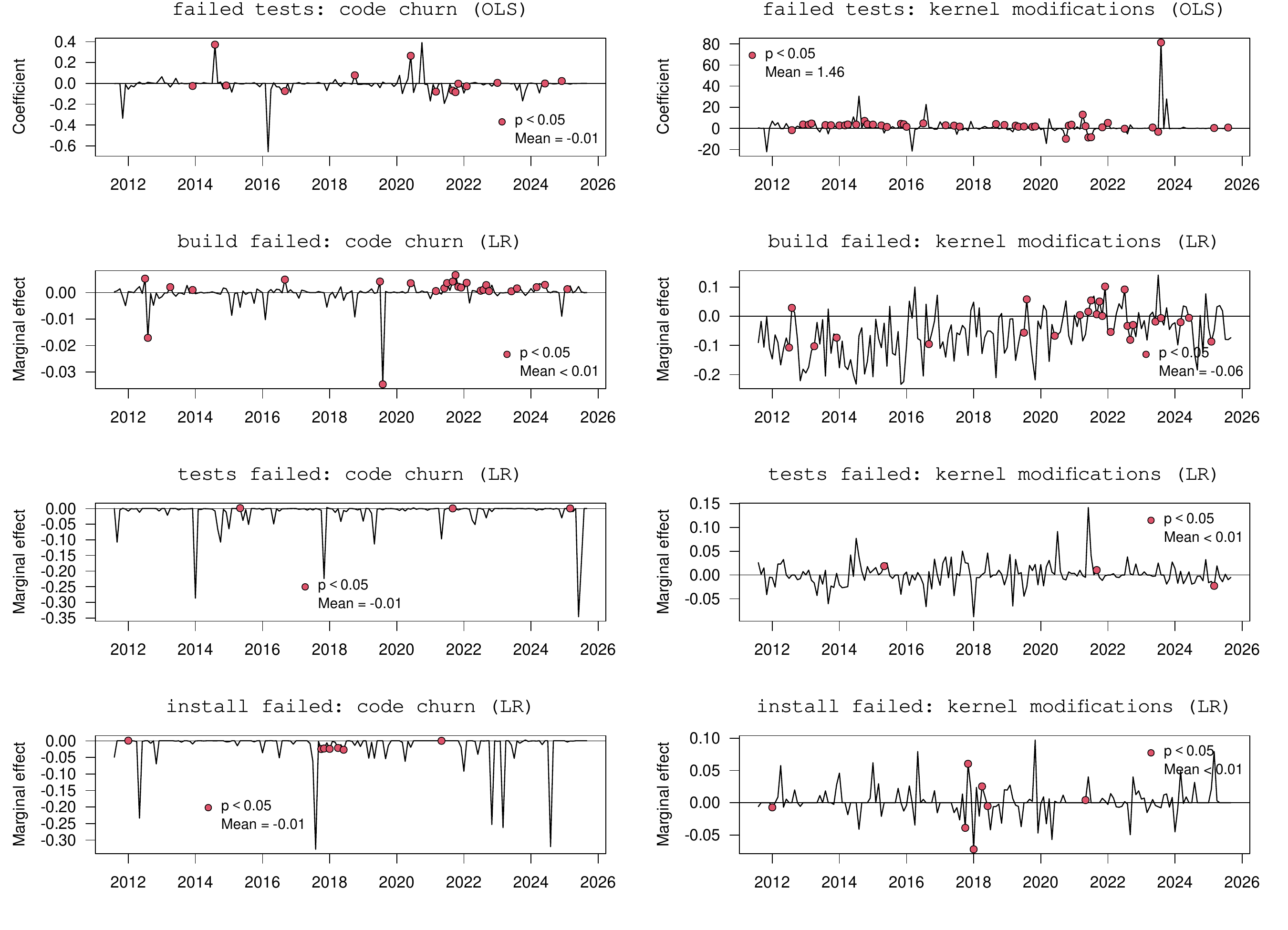}
\caption{Regression Results for the \texttt{i386} Architecture (separate monthly estimates; coefficients or marginal effects of the coefficients)}
\label{fig: reg i386}
\end{figure*}

\subsection{Failures (RQ.2)}

The right-hand side plot in Fig.~\ref{fig: tests} indicates overall long-run
stability of failed test cases over time both with noticeable periods during
which failures have been more frequent. A~similar observation can be made about
the monthly build failure sums shown in the topmost plot in Fig.~\ref{fig:
  failures}. Though, the fluctuations in the two time series do not match
longitudinally. For the failed test results, the period between circa late-2018
to late-2021 stands out, whereas particularly \texttt{amd64} saw many build
failures between 2018 and 2020, and then again in 2023 and 2024. Whatever the
reasons may be behind these fluctuations, it cannot be concluded that the two
time series would be entirely stable. This conclusion is partially supported
also by the twofold sample-splitting test results shown~in~Table~\ref{tab: kw
  tests}.

\begin{table}[th!b]
\centering
\caption{Kruskal-Wallis Non-Parametric Rank Sum Tests}
\label{tab: kw tests}
\begin{tabular}{llrrcrr}
\toprule
&\qquad& \multicolumn{1}{c}{\texttt{amd64}} && \multicolumn{1}{c}{\texttt{i386}} \\
\cmidrule{3-3}\cmidrule{5-5}
Series & Aggregation & $p$-value && $p$-value \\
\hline
\texttt{Failed test cases} & Monthly mean & $0.095$ && $<0.001$ \\
& Monthly median & $0.152$ && $<0.001$ \\
\texttt{Build failed} & Monthly sums & $0.153$ && ~$\phantom{<}0.029$ \\
\texttt{Tests failed} & Monthly sums & $0.080$ && ~$\phantom{<}0.341$ \\
\texttt{Install failed} & Monthly sums & $0.037$ && ~$\phantom{<}0.090$ \\
\bottomrule
\end{tabular}
\end{table}

If the conventional 95\% confidence level is used, the null hypotheses of equal
medians between the two periods are rejected for the monthly mean and median
series for the failed test cases and the monthly sums for the build failures
affecting the \texttt{i386} architecture. Interestingly enough, the null
hypotheses remain in force for \texttt{amd64}, although the
\texttt{amd64}-specific fluctuations are visually more pronounced. A null
hypothesis is also rejected for the installation failures for
\texttt{amd64}. When taking a look at the three series in Fig.~\ref{fig:
  failures}, however, a conclusion based on visual inspection seems better. In
other words, the test suite and install failures can be concluded to be
relatively stable overall despite the fluctuations, whereas a reverse conclusion
can be said to apply to the test and build~failures. Although reporting of test
results can be omitted for brevity, all of the four failure series are still
stationary, meaning that particularly their means remain stable over the whole
period observed. This point provides a statistical time series ground also for
the regression results presented next.

\subsection{Statistical Effects (RQ.3)}

The unaggregated monthly regression estimates are summarized in Fig.~\ref{fig:
  reg amd64} and Fig.~\ref{fig: reg i386} for the \texttt{amd64} and
\texttt{i386} architectures, respectively. In each plot shown in the figures,
the effects of the code churn and kernel modification metrics are shown on the
$y$-axes, whereas the $x$-axes denote the corresponding months used to
separately estimate the OLS and LR regressions. When keeping the points made in
Subsection~\ref{subsec: methods} in mind, it can be started by concluding that
the overall effects are only modest---or even small. This conclusion becomes
clear by taking a look at the means of the effects across the regressions, as
reported in a legend shown in each plot.

The failed test time series seem to be an exception to the conclusion. In
particular, the kernel modification dummy variable shows notable effects upon
the unaggregated test failure amounts. When compared to userland modifications,
one or more preceding commits having touched files in the \texttt{sys} directory
tends to increase the amount of failed \texttt{amd64} test cases by nearly four
tests, all other things being constant. The effect is visible also for
\texttt{i386}, although, as seen from the top-right plot in Fig.~\ref{fig: reg
  i386}, the averaged effect is largely due to a significant effect during a
single month in late 2023. When taking a closer look at the individual plots, a
similar conclusion applies also more generally: the effects may be small on
average, but there are occasionally notable effects during some particular
months. Thus, again, the conclusion about fluctuations apply also to the
regression estimates.

A further point to make is about the signs of the effects. As seen from the
right-hand side plots on the second rows in Figs.~\ref{fig: reg amd64} and
\ref{fig: reg i386}, kernel modifications have tended to decrease the amounts of
build failures. Although various explanations may be possible, one potential
contemplation is that modifying kernel code does not often require modifications
to build scripts, configuration files, and related elements. In fact, the test
suite itself could be seen to imply frequent build failures because each test
case addition requires also modifying bookkeeping material about files present
in an installed NetBSD operating system. Furthermore, regarding the signs, it
can be also seen and concluded that the code churn metric exhibits negative
effects in almost all regressions for the test suite execution and install
failures. This point aligns with the previous speculation about different
sections in the NetBSD's source code tree causing different failures. Not all
failures are common.

\begin{figure}[th!b]
\centering \includegraphics[width=\linewidth,
  height=3cm]{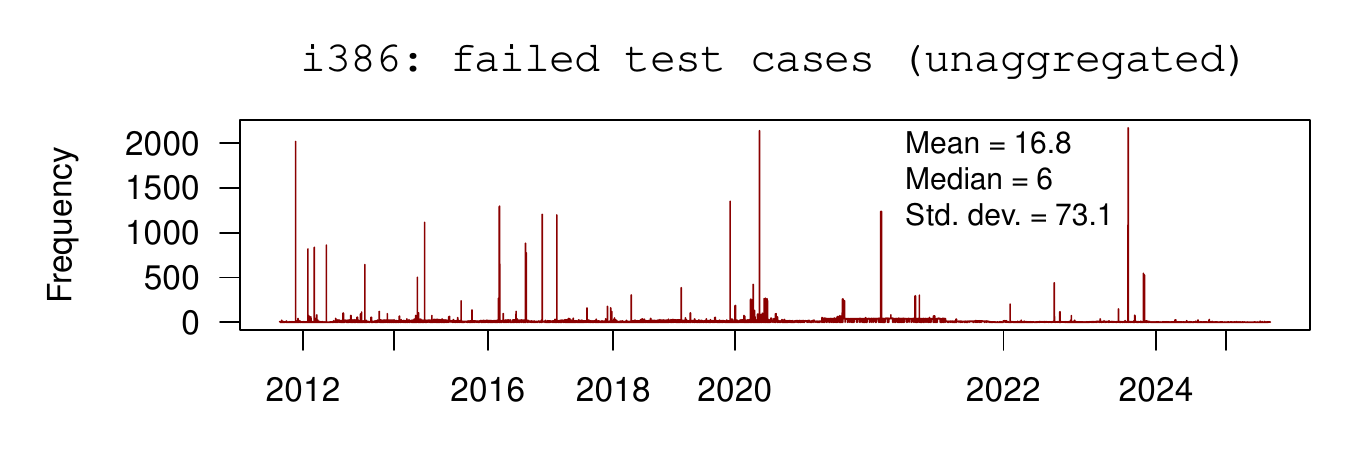}
\caption{An Example of Unaggregated Time Series}
\label{fig: unaggregated example}
\end{figure}

\begin{table}[th!b]
\centering
\caption{OLS and LR Estimates With Pooled Unaggregated Data}
\label{tab: pooled}
\begin{tabular}{llrrrr}
\toprule
& Effect & \texttt{amd64} & \texttt{i386} \\
\hline
Code churn: \\
~\texttt{Failed test cases} & Coef.~(OLS) & $<0.001$ & $<0.001$ \\
~\texttt{Build failed} & Mar.~eff.~(LR) & $<0.001$ & $<0.001$ \\
~\texttt{Tests failed} & Mar.~eff.~(LR) & $<0.001$ & $<0.001$ \\
~\texttt{Install failed} & Mar.~eff.~(LR) & $<0.001$ & $<0.001$ \\
~\\
Kernel modifications: \\
~\texttt{Failed test cases} & Coef.~(OLS) & ~$\phantom{<}5.711$ & $\phantom{<}-0.010$ \\
~\texttt{Build failed} & Mar.~eff.~(LR) & $\phantom{<}-0.088$ & $\phantom{<}-0.027$ \\
~\texttt{Tests failed} & Mar.~eff.~(LR) & $\phantom{<}0.012$ & $\phantom{<-}0.012$ \\
~\texttt{Install failed} & Mar.~eff.~(LR) & $\phantom{<}-0.002$ & $\phantom{<}-0.007$ \\
\bottomrule
\end{tabular}
\end{table}

Finally, to return to the conclusion about modest or even small effects on
average, a few remarks can be made about the pooled regression estimates. Before
continuing, it should be remarked that the pooled unaggregated data yields
irregular time series due to the unequal sample sizes in the monthly archival
files. If \texttt{i386} is taken as an example, the fluctuations in the earlier
Fig.~\ref{fig: monthly n} are seen in a much longer period of observations
between 2020 and 2022 in Fig.~\ref{fig: unaggregated example}, which also
demonstrates the count data characteristics of the pooled data. With these
remarks in mind, it can be seen from Table~\ref{tab: pooled} that the only
notable effect belongs to the kernel modifications dummy variable upon the
failed test cases in the \texttt{amd64} architecture. All other effects are more
or less close to zero. Also the negative binomial regression noted in
Subsection~\ref{subsec: methods} yields coefficients close to zero for both
architectures, including for the kernel modifications. The estimated dispersion
parameters are as large as $227$ and $241$ for the two
architectures. Furthermore, the AR(1) terms yield coefficients with values
$0.784$ and $0.928$. Thus, the pooled OLS estimates in Table~\ref{tab: pooled}
cannot be seen as robust. All in all, these additional checks can be interpreted
to signify that the effects of code churn and kernel modifications are small or
even non-existent on average---yet they are also longitudinally inconsistent,
meaning that there are still some particular short periods during which effects
are visible and likely not noise.

\section{Conclusion}\label{sec: conclusion}

The answers reached are summarized in Table~\ref{tab: results} within which the
symbol \fullcircle~denotes a positive answer with a reasonable confidence, the
symbol \partialcircle~a ``partially'' positive answer with tentative evidence
and interpretation, and the symbol \whitecircle~a negative answer to a
reasonable degree of evidence and interpretation confidence. This categorization
warrants a brief discussion in relation to so-called negative and positive
results.

\begin{table}[t!]
\centering
\caption{Answers to the Research Questions}
\label{tab: results}
\begin{tabularx}{\linewidth}{lcX}
\toprule
Question & Answer & Explanation \\
\hline
RQ.1 & \fullcircle & The test suite has continuously grown. \\
\hline
RQ.2\one & \partialcircle & Overall, the amounts of failed test cases have been stable, but the time series indicate also fluctuations during which failures have been more frequent. \\
\cmidrule{3-3}
RQ.2\two & \partialcircle & Although overall stability is present, there have also been occasional periods marked with frequent build failures, and these failure periods also differ between the \texttt{amd64} and \texttt{i386} architectures. \\
\cmidrule{3-3}
RQ.2\three & \fullcircle & There has been overall stability with some short-lived bursts. When compared to the answer to RQ.2\two, the fluctuations have been short and affected particularly \texttt{i386}, possibly due to more frequent runs of the test suite for \texttt{i386}. \\
\cmidrule{3-3}
RQ.2\four & \fullcircle & The answer is analogous to the one for RQ.2\three. \\
\hline
RQ.3\one & \partialcircle & \onecase~Code churn and \twocase~kernel modifications do not  \\
RQ.3\two & \partialcircle & have consistent effects upon the four failure cases on average. Only the kernel modifications show \\
RQ.3\three & \partialcircle & notable effects upon test failures. Yet, there are \\
RQ.3\four & \partialcircle & periods during which other effects are visible too. \\
\bottomrule
\end{tabularx}
\end{table}

Negative results---understood generally as results that do not support prior
expectations---have been praised also in software
engineering~\cite{Paige17}. However, there are reasons why the concept is poor
for describing the results reached and presented. To put aside the issue whether
it even makes sense to dichotomously frame results as negative (positive) with
respect to some prior positive (negative) expectations~(cf.~\cite{McShane17}),
the exploratory nature of the paper prevents characterizing the results as
negative. In other words, none of the research questions that were specified in
Section~\ref{sec: background and related work} came with solid prior
expectations seeking either an empirical confirmation or an empirical
falsification. Rather, the many \partialcircle~symbols in Table~\ref{tab:
  results} would support an alternative term of inclusive
results~(cf.~\cite{Borji18}). This point applies particularly for the answers
given to RQ.3; the effects are small or even non-existent on average but the
conclusion cannot be generalized to all periods observed. An analogous point
applies with respect to the fluctuations affecting the answers to the question
RQ.2.

\section{Limitations}\label{sec: limitations}

Regarding limitations, the synchronization issues noted in
Subsection~\ref{subsec: aggregation} are a threat to internal validity. Also the
statistical issues noted in Subsection~\ref{subsec: methods} can be mentioned in
this regard. Although there are arguably no right or wrong answers, the
operationalization of code churn (see Subsection~\ref{subsec: metrics}) could be
seen as a small threat to construct validity.

As a case study, external validity too is an inevitable
limitation. However---because test suites differ substantially between projects,
it seems sensible to expect that generalizability remains an issue also in
further research. In contrast, the synchronization issues would seem more
plausible to address in further research. It should be noted that the issue is
not really about merging the monthly archival files into a single file or a
database entry---such a task seems almost trivial. Rather, the issue is more
about how to do sensible time series analysis when data is irregular and should
be broken into smaller subsets. The question could be extended to bisecting of
failures that is a common task and functionality in automated testing
frameworks~\cite{Ruohonen19RSDA}. Also the NetBSD's testing framework does
bisecting to buggy commits but it is unclear how robustly.

\section{Further Research}\label{sec: further research}

The inconclusive results would provide an opportunity to move beyond the
exploratory ``what'' questions that were noted in Subsection~\ref{subsec: code
  churn and kernel code}. In other words, it would be worthwhile to understand
why there have been fluctuations, what has caused them, and how NetBSD
developers themselves perceive them. Likewise, it would be relevant to know why
kernel modifications may sometimes, but not always, cause more test cause
failures but less build failures, and so forth. Though, there are many ``what''
questions that would warrant further research too; among other things, the paper
did not examine what is actually being tested in NetBSD. The point about
potential stabilization or even saturation made in Subsection~\ref{subsec:
  growth} would offer a good research question for further research in this
regard. There is also an alternative hypothesis in this regard.

Based on anecdotal evidence from NetBSD developers, the pace of development has
allegedly slowed down in recent years. It could well be that this supposed
slowdown rather than stabilization explains the trends in Fig.~\ref{fig:
  tests}. This hypothesis conveys a broader methodological point: the evolution
of test frameworks and test cases should be studied in conjunction of the
evolution of what is being tested. Coverage evolves too.

\subsection*{Acknowledgements}

The authors thank NetBSD developer Andreas Gustafsson for an additional review
and helpful comments.

\balance
\bibliographystyle{IEEEtran}


\end{document}